\documentclass[conference,compsoc]{IEEEtran}
\IEEEoverridecommandlockouts

\usepackage{cite}
\usepackage{ulem}
\usepackage{comment}
\usepackage{amsmath,amssymb,amsfonts}
\usepackage{algorithmic}
\usepackage{graphicx}
\usepackage{textcomp}
\usepackage{enumerate}
\usepackage{enumitem}
\usepackage{xcolor}
\usepackage{soul}
\usepackage{hyperref}
\usepackage[utf8]{inputenc}
\usepackage{multirow}
\def\BibTeX{{\rm B\kern-.05em{\sc i\kern-.025em b}\kern-.08em
    T\kern-.1667em\lower.7ex\hbox{E}\kern-.125emX}}
\newcommand{\mypara}[1]{\smallskip \\ \noindent \textbf{#1: }}

\author{\IEEEauthorblockN{Karen Sowon\IEEEauthorrefmark{1}, Edith Luhanga\IEEEauthorrefmark{2}, Lorrie Faith Cranor\IEEEauthorrefmark{1}, Giulia Fanti\IEEEauthorrefmark{1}, Conrad Tucker\IEEEauthorrefmark{1} and Assane Gueye\IEEEauthorrefmark{2}}
\IEEEauthorblockA{\IEEEauthorrefmark{1}
Carnegie Mellon University}
\IEEEauthorblockA{\IEEEauthorrefmark{2}Carnegie Mellon University-Africa}
}

\begin{document}
\pagenumbering{arabic}

\title{The Role of User-Agent Interactions on\\ Mobile Money Practices in Kenya and Tanzania}
\maketitle

\begin{abstract}
Digital financial services have catalyzed financial inclusion in Africa. Commonly implemented as a mobile wallet service referred to as mobile money (MoMo), the technology provides enormous benefits to its users, some of whom have long been unbanked. While the benefits of mobile money services have largely been documented, the challenges that arise---especially in the interactions between human stakeholders---remain relatively unexplored. In this study, we investigate the practices of mobile money users in their interactions with mobile money agents. We conduct 72 structured interviews in Kenya and Tanzania (n=36 per country). The results show that users and agents design workarounds in response to limitations and challenges that users face within the ecosystem. These include advances or loans from  agents, relying on the user-agent relationships in place of legal identification requirements, and altering the intended transaction execution to improve convenience. Overall, the workarounds modify one or more of what we see as the core components of mobile money: the user, the agent, and the transaction itself.  The workarounds pose new risks and challenges for users and the overall ecosystem. The results suggest a need for rethinking privacy and security of various components of the ecosystem, as well as policy and regulatory controls to safeguard interactions while ensuring the usability of mobile money. 
\end{abstract}


\section{Introduction}
\label{Sec:Intro}
The pervasive adoption of mobile phones in many low-and-middle-income countries (LMICs) has led to the widespread use of mobile phones for delivery of financial services. One such digital financial service is mobile money (MoMo) which makes financial services available on the mobile phone through one or a combination of the following technologies: unstructured supplementary service data (USSD), short message service (SMS), SIM Toolkit (STK), and smartphone apps. The use of SMS, STK, and USSD ensures that the service can be provided on basic phones, which a larger majority in LMICs have access to \cite{silver_johnson_2018}.  

\begin{figure}[htbp]
\includegraphics[width=\linewidth,scale=0.5]{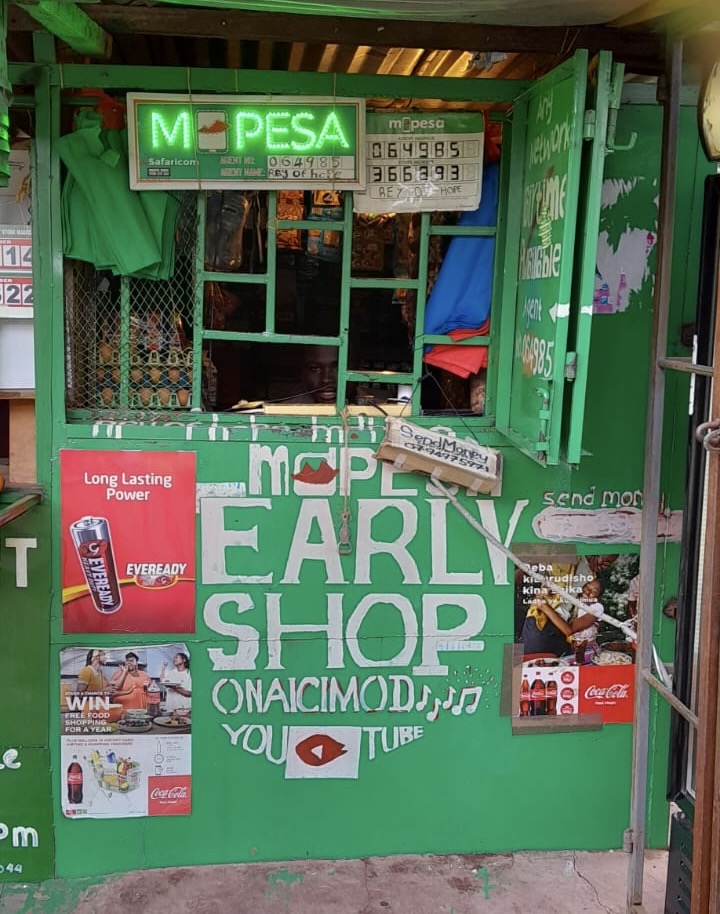}
\caption{An M-PESA shop in Kenya (credit: Karen Sowon)
}
\label{figAgent}
\end{figure}

Mobile money has helped LMICs overcome many traditional banking challenges and has spurred financial inclusion and driven economic growth \cite{demirgucc2022global}. At the center of the definition of financial inclusion is the idea of availability, use, and accessibility, which is measured by the reach of financial services to individuals including the underserved \cite{demirguc2018global}. While there are still close to 1.5 billion people in the world, mostly in Africa, who lack access to basic financial services 
\cite{demirguc2018global} \cite{demirgucc2022global}, MoMo has been instrumental in making financial services more widely available. In 2020, there were approximately 1.2 billion registered mobile money accounts globally, and over 50\% of these were from sub-Saharan Africa (SSA) \cite{andersson2021state}. The use of MoMo contributed to narrowing the gap of un-banked people from 2.5 billion in 2011 down to 1.7 billion people in 2017 \cite{demirguc2018global}. Accessibility is achieved by having an extensive  network of agents (Figure \ref{figAgent}). These are third parties acting on behalf of the MoMo service providers---mostly mobile network operators (MNO)---to facilitate digitization of money. This network is the backbone of MoMo services since it acts as a distribution channel and provides a trusted mechanism for people to transact.
With a far extended reach of about 55 times that of banks and 26 times that of ATMs \cite{jakachira_andersson_2020}, agents act as an alternative to bank branches, and provide ``last-mile access,'' hence overcoming the distance-related barriers of owning and managing an account at a formal financial institution. Agents digitize more than \$715 million per day \cite{awanis2022state}. Overall, MoMo contributes enormously to the economy with transactions representing up to 75\% of the country’s gross domestic product (GDP) in  countries like Kenya, the home of one of the earliest implementations of MoMo, M-PESA \cite{creemers2020five}.

While MoMo agents and users are central actors to the ecosystem, little is known about the intricacies of their interaction and the overall impact this has on the use of MoMo. Although low trust is one of the hindrances to the use of formal financial products, we know from MoMo adoption literature that trust is an important antecedent to its use \cite{noreen2021impact} \cite{okello2020trust} \cite{osakwe2022trust}, considering that one of the first interactions that users will have entails handing over physical cash to the agent \cite{davidson2011driving} \cite{naghavi2019state}. In addition to facilitating cashing-in and cashing-out services (CICO), agents also provide additional services such as on-boarding new users and offering technical assistance \cite{andersson2021state}. Given the essential role that agents play, interactions between users and agents can have serious privacy and security implications for users that are not currently well-understood. We therefore study users' habits qualitatively, to get a rich understanding of the ways that users interact with agents, and how this can affect the privacy and security of the MoMo ecosystem at large. Our study investigates these research questions:
\begin{itemize}
\item RQ1a: What practices do the relationship between agents and users facilitate and why do users engage in these practices? 
\item RQ1b: What privacy and security challenges contribute to or arise from these practices? 
\item RQ2: How do users navigate these privacy and security challenges? 
\item RQ3: What are the implications and challenges of MoMo user-agent interactions and practices? 
\end{itemize}
Our study presents the results of an exploratory qualitative study to understand the practices of MoMo users in their interaction with agents in two East African countries with high MoMo adoption rates: Kenya and Tanzania with 73.12 million and 40.9 million registered accounts respectively as of 2022  \cite{tcraquarterly2022}\cite{kenyamobilemoney}. 
We conducted structured interviews between July and September 2022 with MoMo users between the ages of 18 and 65. We used inductive thematic analysis to synthesize the data.

Due to rapid population growth in LMICs, the demand for MoMo services may remain high for a long time, especially among the millions who are marginalized \cite{awanis2022state}. Thus, it is crucial to understand the interaction between users and agents to safeguard the opportunities for financial inclusion.

Our findings in this work highlight three important, previously-undocumented findings related to the interaction between clients and MoMo agents:
\begin{itemize} %[align=left]
    \item First, we find that users have a range of reasons for choosing a particular agent over another. These reasons are often related to trust and/or convenience, as well as lax know-your-customer (KYC) practices. The reasons for choosing a particular agent impact how users interact with that agent.
    \item Second, we identify and categorize a number of challenges that users encounter in their interactions with MoMo systems in general, and agents in particular. These challenges relate to complying with KYC requirements, privacy and security concerns surrounding information sharing with agents, and interface usability. Our data help to explain why these challenges arise; for instance, many challenges arise due to cost and a lack of sufficiently convenient access to agents. 
    \item Third, our results illustrate a wide range of workarounds or practices that both users and agents engage in to circumvent these challenges. These workarounds often introduce increased  risk to users, thwarting the very measures that MoMo providers have put in place to facilitate privacy and security. Some of these workarounds involve changing the location or intended process of a transaction, while others actually involve agents breaking KYC protocol, as well as offering services that are not officially supported by the MoMo provider such as loans and advance transactions. 

\end{itemize}

Overall, our results suggest that agents---thanks to their willingness to flexibly navigate or modify intended MoMo procedures---play a critical role in making MoMo accessible and convenient for end users. At the same time, the challenges users experience allow us to offer concrete recommendations for MNOs, researchers, and regulators. 

First, we identify a need to improve the MoMo user experience for confirming and reverting transactions. 
Second, our results suggest that users are often hesitant to share sensitive data such as transaction amounts (especially if very large or very small) with agents. We recommend further research to understand how and how often user privacy concerns are affecting MoMo usage, as well research on solutions that would limit agents' data access during interactions. 
Third, we suggest developing registration and KYC mechanisms for users who lack formal ID. Many of the workarounds we observe are necessitated by regulations requiring SIM cards to be registered to a formal ID, which causes significant obstacles to those without ID. The recommendations are not entirely prescriptive because solutions to improve MoMo would need to be accompanied by studies to assess their feasibility and usability.

\section{Background on Mobile Money}
\label {Sec:background}
MoMo differs from other mobile-based financial services, such as mobile banking or mobile payments in developed countries, in three key ways \cite{nan2019mobile}. First, MoMo initiatives are specifically designed to cater to the financial needs of individuals who are excluded from formal financial services (the un-banked). Second, users do not need a formal bank account since the user's phone acts as an account. Finally, MoMo relies on widely-available transactional agents. 
These features make MoMo an effective digital tool to extend basic financial services to individuals who would otherwise be financially excluded. 
 
\subsection{Understanding mobile money operations} 
Assuming the role of a financial service provider, the MoMo providers work with agents who are mostly small businesses within the community, and therefore extend financial services in a non-intimidating and familiar environment \cite{mckee2015doing}. With a MoMo account, individuals can use their mobile phones to execute financial transactions such as cashing in, cashing out, or person-to-person (P2P) transfers (Figures 2-4). They may also pay their bills and pay for other goods and services \cite{osakwe2022trust}. However, loan facilities are not provided through agents. The mobile wallet is often protected by a PIN, and most MoMo services adopt some form of additional authentication to identify users. Both Kenya and Tanzania, like most African countries, mandated SIM card registration with some proof of identity \cite{theodorou2021access}. Both countries require national IDs for this process, with Tanzania also collecting biometrics. In addition to these SIM registration requirements to identify users,  Kenya requires MoMo users to show their ID card to the agent when they transact. These authentication or KYC procedures are meant to protect against fraud and money laundering. 

\begin{figure}[t]
\includegraphics[width=\linewidth]{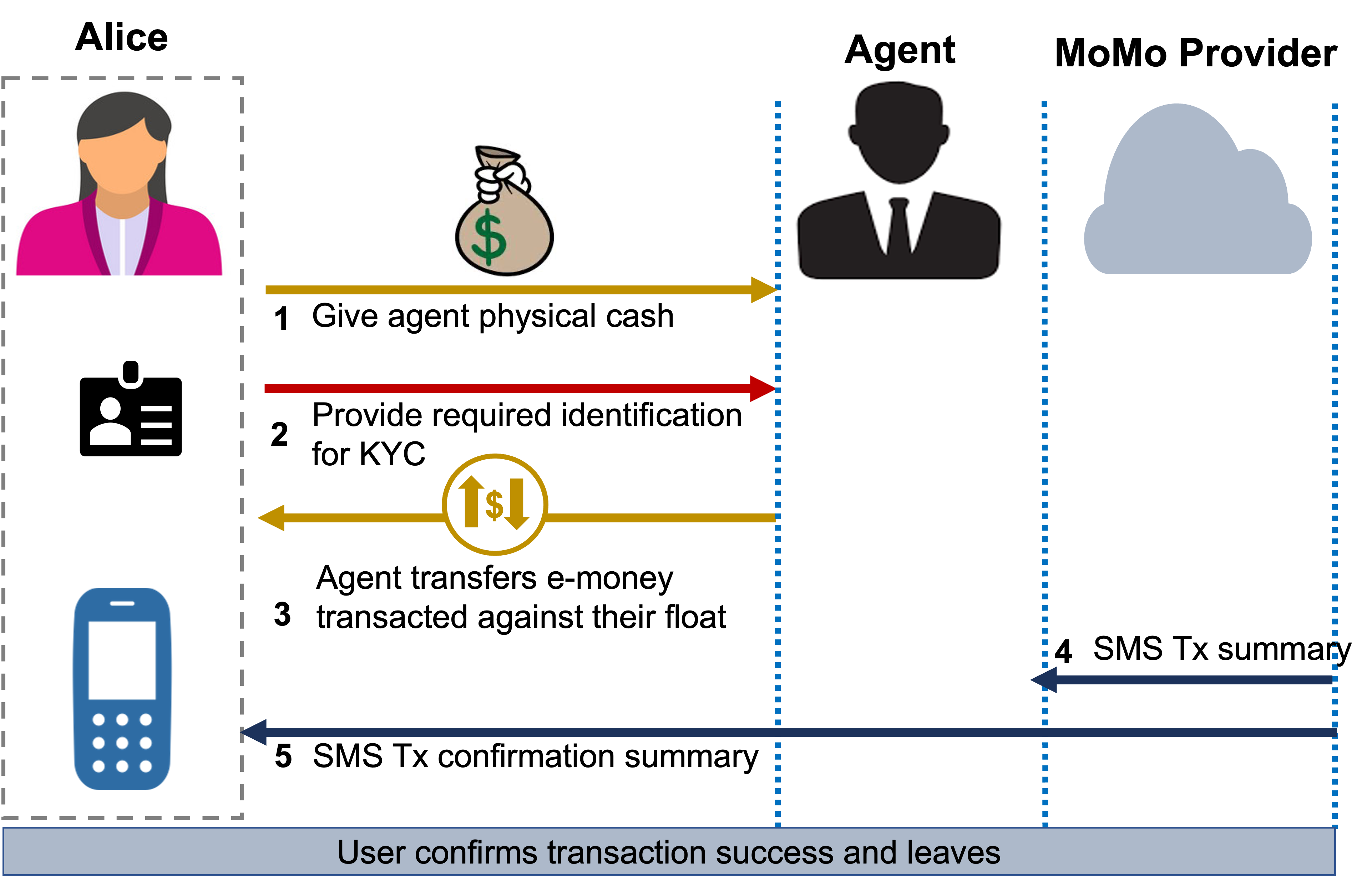}
\caption{Standard scenario for Cashing-In (Deposit)}
\label{figDeposit}
\end{figure}

\begin{figure}[t]
\includegraphics[width=\linewidth]{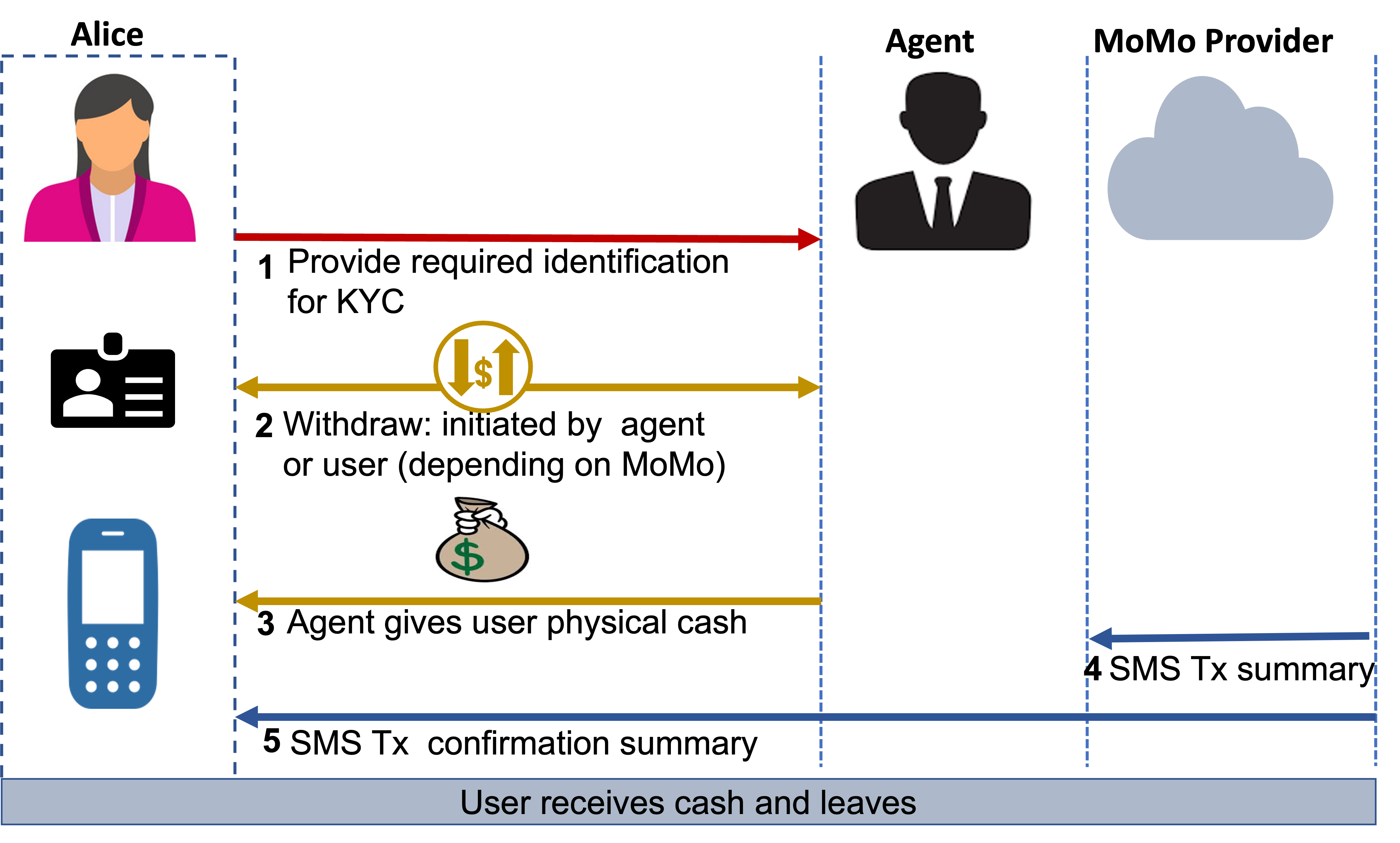}
\caption{Standard scenario for Cashing-out (Withdraw)}
\label{figWithdraw}
\end{figure}

\begin{figure}[t]
\includegraphics[width=\linewidth]{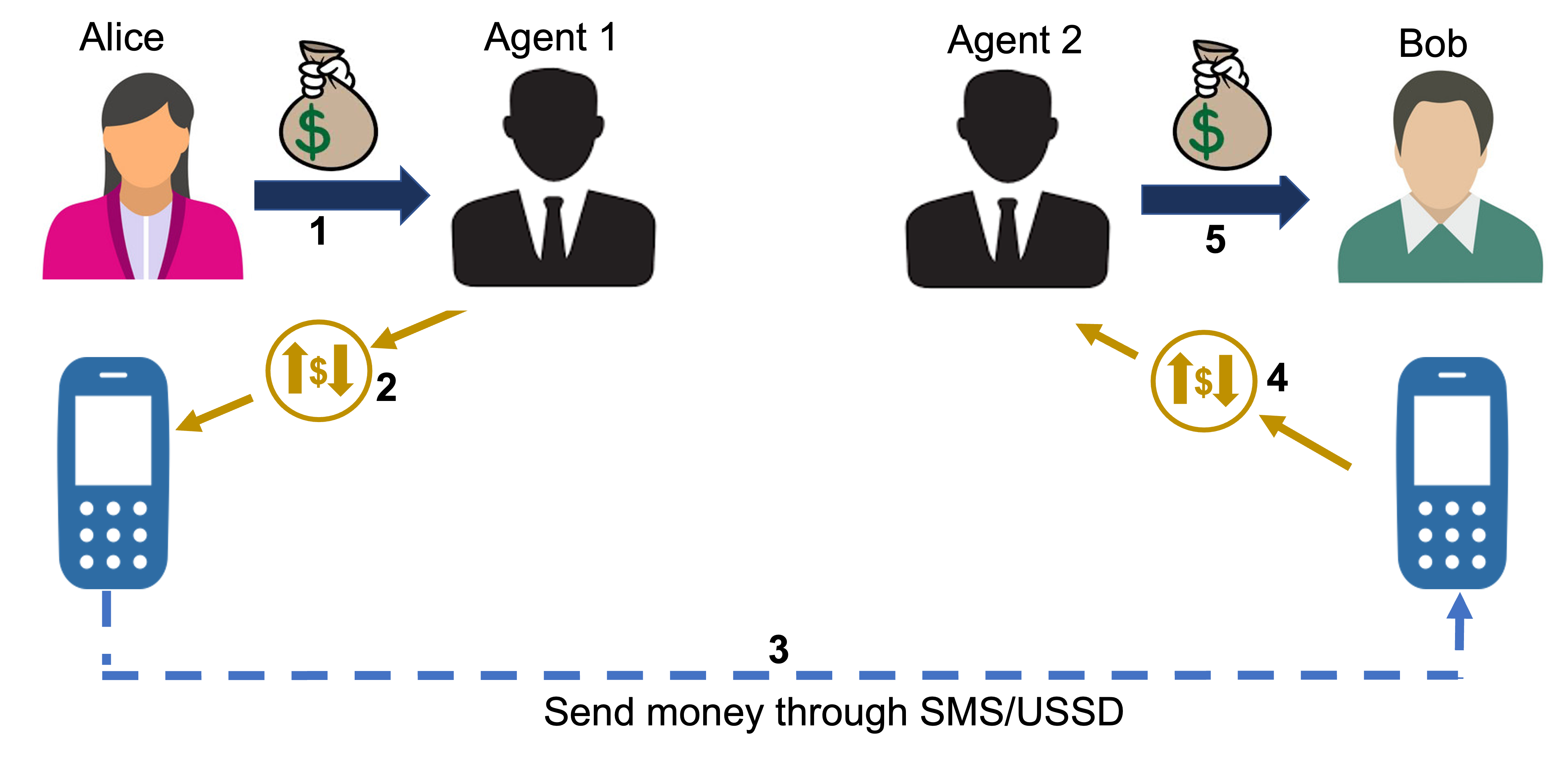}
\caption{Standard scenario for a P2P transaction}
\label{P2P}
\end{figure}

\subsubsection{Transaction life cycles}
\label{sec:floatdef}
The MoMo operators provide agent outlets with special e-wallets that have higher maximum account balances than user wallets.
These e-wallets provide an e-money store \textit{float} against which the agent balances their customers' cash-in/cash-out (CICO) transactions. If the agent conducts too many cash-in transactions, they will eventually exhaust their e-float, and if they perform too many cash-out transactions, they will run out of cash. To restore balance, the agent needs to convert excess e-float into cash or vice versa, at a higher cash distribution entity such as a bank or a super agent. Here we illustrate the three main types of transactions that a MoMo user can accomplish. 
\mypara{Cashing-in (Depositing money, also known as topping up)}
Suppose Alice needs some money in her mobile wallet  (Figure \ref{figDeposit}). Alice goes to the MoMo agent and hands over her physical cash in exchange for e-money. As the agent transfers the e-money equivalent to Alice, the agent's float will reduce.
\mypara{Cashing-out (Withdrawing money)}
Suppose Alice has some e-money in her wallet and needs physical cash (Figure \ref{figWithdraw}). Alice goes to the MoMo agent where the cash-out transaction will be initiated either by Alice or by the agent. The transaction reduces Alice's balance but increases the agent's float. The agent  hands  the cash equivalent to Alice.
\mypara{Person-to-person transactions (P2P)}Suppose Alice wants to send money to Bob  (Figure \ref{P2P}). 
Alice goes to the MoMo agent (Agent 1) and tops up her MoMo mobile account (assuming she does not already have any balance in her MoMo wallet). Alice then accesses the MoMo service on her phone.  
Alice sends the money to Bob by entering the amount, and Bob's mobile number that acts as his account, and (in some cases) confirms by entering her PIN. Bob can go to the nearest MoMo agent (Agent 2) and withdraw the money from his account using the previously described cash-out process.

\section{Related Work}
\label{Sec:related work}
We discuss two strands of prior work related to this study: adoption and use of MoMo, and privacy and security of digital financial services.  

\subsection{Adoption of mobile money}
Some studies have identified the barriers to the use of MoMo, most of which mirror the top barriers to account ownership at a formal financial institution. These include cost, unreliable mobile networks and power grids, preference for cash, lack of documentation for KYC requirements, and low literacy---including digital literacy skills \cite{awanis2022state} \cite{demirgucc2022global}. These studies also suggest that although perceived trust of MoMo contributes to its use among users, lack of trust is one of the main barriers cited by non-users \cite{akinyemi2020determinants} \cite{awanis2022state}.

Similarly, technology adoption studies (e.g.,\cite{baganzi2017examining},\cite{abdul2019customers}, \cite{okello2020trust}, \cite{noreen2021impact}, \cite{osakwe2022trust})  have investigated the factors that predict uptake and use of MoMo. Most of these studies are  quantitative investigations that aim to establish antecedents to adoption. The findings in these studies unanimously emphasize the importance of trust as a significant predictor of adoption. Osakwe et al., \cite{osakwe2022trust} further explore factors that can increase trust in MoMo and offer four specific  institutional measures that can support initial trust building. One of the factors is perceived reputation of both the MoMo agents and the MoMo provider. While these studies have been useful in elucidating adoption, none of them explore the interactions that arise from trust between the agents and users, either pre-adoption or post-adoption
\cite{benbasat2005trust} \cite{sowon2020trust}. This trust often influences continued use.

Post adoption, users tend to develop technology coping strategies in response to technology challenges. Other researchers have studied this in relation to technologies like mobile apps, mobile health, and tablets \cite{inal2019users} \cite{sowon2022information} \cite{zamani2021accommodating}. In general, these studies show that users can cope with technology by abandoning it (the flight response) or accommodating if (the fight response). The concept of workarounds that we use to describe alternative workflows in this study can be considered a fight response.
  
\subsection{Mobile financial system privacy and security}
The increased uptake of MoMo has been associated with an increase in fraud that takes advantage of security vulnerabilities in MoMo technologies \cite{reaves2017mo}. Some of the fraud cases center on the role of agents. In 2020, Uganda suspended its MoMo transactions on its network after hackers breached the payment system, resulting in a loss of over \$3 million  \cite{kafeero_2020}. The attack was perpetrated by fraudsters who worked in collaboration with MoMo agents. Researchers have also analyzed the security of MoMo in Uganda and identified the potential sources of security risks \cite{ali2020evaluation}. 
While our study touches on the role of fraud in MoMo users' decision-making,  it is only a small component of our study. More broadly, we want to understand how user beliefs and perceptions affect their interactions with agents and the MoMo system at large, whether those perceptions relate to fraud or not.

Similarly, several research studies have explored MoMo privacy issues, reviewing privacy policies and relevant regulations   \cite{bowers2017regulators} \cite{harris2012privacy}. MoMo is a data-rich environment, and prior work has noted various aspects of the transaction process that may raise privacy concerns \cite{makulilo2015privacy}. These include the identification of subscribers through SIM registration, KYC performed by agents, and the multiple roles that agents play that make them data depositors. The interaction between various stakeholders also introduces layers of complexity for privacy and security in MoMo environments \cite{mogaji2022dark}. Other studies have focused on privacy concerns with related services such as mobile loan apps \cite{munyendo2022desperate} and some have explored the role of human intermediaries on access to digital services. For example, in Kenya,  cybercafe owners sometimes performed services for their patrons such as logging in~\cite{munyendoeighty}.  While addressing privacy, these studies do not specifically evaluate the user-agent interaction. 

In the work most closely related to ours, Mogaji and Nguyen \cite{mogaji2022dark} explore the ``dark side" of MoMo interactions between stakeholders in Nigeria. 
Their study touches on user-agent relationships, from which they observe several similar patterns to those in our work, including unofficial fees from agents and third-party data sharing. 
The most important difference between their work and ours is that we study the user-agent relationship in depth from the perspective of users in Kenya and Tanzania, whereas their work considers multiple stakeholders (including users and agents), and focuses mainly on the downsides of MoMo for financially-vulnerable customers in Nigeria.
As a result, their study does not uncover the same breadth or depth of workarounds or practices that we observe, nor does it probe the underlying causes for these workarounds in as much depth, if at all. 
\section{Methods}
\label {Sec:methodology}
Given the under-explored nature of the practices of mobile money users in prior work, we adopted an exploratory approach using qualitative methods. The data collection was completed between July and September 2022 with the help of a contractor identified through a formal bidding process. 
The interview we present here was immediately preceded by another interview during the same session that we report on in a separate paper. 

\subsection {Instrument design and pilots}
All tools, data collection procedures, and recruitment and screening material were developed by two lead researchers, natives of Kenya and Tanzania, with feedback from a team of co-authors with complementary expertise. This was useful in providing  interpretation of contextual nuances while developing the data collection instruments. We opted for structured interviews (Appendix \ref{Appendix_IDI}) to maintain consistency across the contract research assistants working in the two countries. The user-agent interview centered on the following topics: 
\begin{itemize}
    \item practices of MoMo users through the customer journey,
    \item perceptions of and experiences with agents, and
    \item security and privacy concerns in the use of MoMo.
\end{itemize}  

The questions were informed by prior studies that suggest the influence of trust in user behaviour and  practices such as agent selection \cite{chamboko2021role}, information disclosure \cite{yisa2023investigating}, and security delegation \cite{forget2016or}.

We conducted two iterations of cognitive walkthroughs with an independent expert and modified the questions for clarity after each iteration. We then completed three initial pilots with participants similar to  those in the target population, made further modifications, and completed two more pilots. We do not include pilot interview data in the results presented here. The two native researchers worked with native student assistants to translate the interview guides to Kiswahili, a language spoken in both Kenya and Tanzania, albeit with different regional nuances. 

\subsection{Participant sample and recruitment}
 The research assistants recruited 36 participants within each country. The respondent population was MoMo users in Kenya and Tanzania over the age of 18 years. 
We use stratified purposive sampling to achieve heterogeneity of the sample by gender, location (rural and urban), and age (Table \ref{table:demographics}). Although we planned in advance to conduct at least 36 interviews, we stopped there due to reaching saturation.
\begin {table} [htbp]
\caption{Number of participants per category}
\centering
\begin {tabular} {|p{1cm}|p{1.3cm}|p{1.2cm}|p{1.5cm}|}
\hline
& &Kenya & Tanzania \\
\hline
Age& 18-25& 13& 9 \\
& 26-39& 16& 14 \\
& 40-49& 3& 5\\
& 50-65& 4& 8\\
\hline
Gender& Female& 16& 18\\
& Male& 20& 18\\
\hline
Location& Urban& 18& 17\\
& Rural& 18& 19\\
\hline
\end{tabular}
\label{table:demographics}
\end{table}

\subsection {Data collection}
Each user-agent interview lasted approximately 20-40 minutes, with the total interview time for the two interviews averaging 75 minutes. All interviews in Kenya were face-to-face: 12 in English, 10 in Kiswahili, and 14 using a mix.\footnote{In Kenya, people often speak by mixing English and Kiswahili, and sometimes this mix morphs to `sheng' which is a Kiswahili and English-based slang} In Tanzania, all interviews  were over the phone, in Kiswahili, with the exception of one interview that used a mix. The different data collection methods in the two countries was necessitated by complex national protocol modification processes that were beyond the researchers' control. Research has shown that while different modes of interview data collection may present logistical variations, there is little to no impact to validity of interview data \cite{oates2022audio}. We asked the security- and privacy-related questions towards the end of the interview to minimize bias in how users may have perceived their interactions and actions with the agent. 

\subsection{Data analysis}
The contractor translated the Kiswahili transcripts into English, and submitted all transcripts to the researchers. The native lead researchers conducted an inductive thematic analysis of the data using NVivo qualitative data analysis software \cite{braun2006using}. This began with one researcher checking all transcripts to ensure de-identification and renaming them using pseudonyms: KE01-KE36 and TZ01-TZ36 for participants from Kenya and Tanzania respectively. Following this, the two researchers independently read a 20\% sample of the transcripts and jointly developed an initial codebook, which they used to independently code a second  20\% sample of the transcripts. They met to discuss and resolve any conflicts. One of the two researchers then coded the remaining transcripts, with the second researcher spot checking random samples of the work to ensure consistency; the two met to discuss whenever there were coding differences. Coding differences were largely around code definitions and the importance of particular data points, which we resolved by adding clear code descriptions with sample quotes, and revisiting the study questions respectively. When there was a codebook change, the researcher re-coded the full set of excerpts with the new scheme. We required three coding iterations to be sure that the codes had been applied consistently. The codes (Appendix \ref{Appendix_Codes}) were also discussed with the entire  research team as they emerged. We organized the resulting codes into final themes and sub-themes  (Appendix \ref{Appendix_ThemeEmergence} before selecting useful excerpts for reporting purposes.  

\subsection{Ethical considerations}
This study was approved at the national country levels, and at our institution's IRB. See Appendix \ref{appendix:approvals}. We collected participants' phone numbers for the purpose of compensation via mobile money. However, these were stored separately from the interview transcripts. All participants also provided informed consent to be interviewed and audio-recorded, and were compensated in amounts approved by the local ethics committee: US \$3.40 in Kenya, and US \$10 and \$5 in Tanzania for urban and rural participants respectively. These amounts were for both interviews together, and the difference in Tanzania was based on estimates for hourly wages in urban areas, and daily work output in rural areas in Tanzania. Recordings were stored in password-protected files that were shared between the researchers and the country leads who were responsible for overseeing all data collection activities for their respective countries. Transcripts were anonymized and de-identified before analysis. 

\subsection{Limitations}
The study has some methodological limitations. Given the qualitative nature of the study and the small sample size, our results may not be generalizable to entire populations of MoMo users. Some nuances may also have been lost in translation of the interviews between  languages, which was done by native research assistants working with the native researchers.
The cross-sectional design of the study also poses additional limitations. Regulators, governments and policy-makers are regularly introducing or changing aspects of this ecosystem and we expect that user  practices will also continuously adapt to this changing environment. 
While we did our best to minimize bias in the design of the study, our results may suffer from self-reporting and social desirability bias if participants felt the urge to report ``proper'' behaviour. 
Finally, in an effort to ensure consistency across several assistants and within the two countries, the decision to use structured interviews may have limited the richness of the data that we could collect.   
\section{Results}
\label {Sec:Results}

We present our results for each of the three research questions posed in the introduction. In most cases we found similar trends in Tanzania and Kenya so we report merged findings. We highlight areas where we observed different trends in the two countries. While we often provide the number of responses corresponding to a particular code to give readers a sense of the frequency of each theme in our interviews, we caution that this is a qualitative study and these numbers should not be interpreted as representative of frequencies of beliefs and behaviors in the population.  
\begin {table} [htbp]
\caption{A summary of workarounds by users and agents}
\centering
\begin {tabular} {|p{3cm}|p{2.7cm}|p{1.8cm}|}
\hline
\multicolumn{2}{|l|}{\textbf{Transaction Execution Workarounds}}& \textbf{Changing Component}\\
\hline
\multirow[c]{2}{*}{\textbf{Alternative execution}}& Agent executes transfer& Agent \\ \cline{2-3} & Transact at a distance & User \& Agent\\
\hline
\multirow[c]{2}{*}{\textbf{Alternative services}}& Advance transactions& \multirow[c]{2}{*}{Agent} \\ \cline{2-2} & Credit/Loan services & \\
\hline
\multirow[c]{3}{3cm}{\textbf{Change transaction characteristics}}& Location& \multirow[c]{3}{*}{Transaction} \\ \cline{2-2} & Size & \\ \cline{2-2} & Channel & \\
\hline
\multicolumn{2}{|l|}{\textbf{KYC Workarounds}}& \\
\hline
\multirow[c]{2}{*}{\textbf{Agent-adjusted KYC}}& Relationship-based& User \& Agent \\ \cline{2-3} & Agent complaisance & Agent\\
\hline
\multirow[c]{3}{3cm}{\textbf{Alternative identity proofing}}& Third-party ID& \multirow[c]{3}{*}{User} \\ \cline{2-2} & Reduced KYC & \\ \cline{2-2} & Human recommender & \\
\hline
\multicolumn{2}{|l|}{\textbf{Confirmation Workarounds}}& \\
\hline
\multirow[c]{2}{*}{\textbf{Alternative confirmation}}& Agent confirmation& User \& Agent \\ \cline{2-3} & Recipient confirmation & User\\
\hline
\multicolumn{2}{|l|}{\textbf{Transaction Reversal Workarounds}}& \\
\hline
\textbf{Alternative reversal}& Agent-assisted reversal& Agent\\
\hline
 \end{tabular}
\label{table:tab1workaroundsContinuum}
\end{table}

\subsection{Q1: Practices facilitated by the user-agent relationship, motivations, and privacy/security concerns}
\label{sec:habitsandpractices} 
The practices we observed can be roughly divided into two categories: 1) practices surrounding how mobile money users \textit{choose} an agent, and 2) practices around how mobile money users \textit{use} an agent. 
The latter category was characterized by surprising examples of usage that we term \textit{workarounds}---workflows that are not intended by the MoMo provider, but that clients and agents, either independently or sometimes in a concerted way, work out between themselves. We identify these workarounds (Table \ref{table:tab1workarounds}) according to the stage of the transaction continuum in which they took place:  during the transaction execution, KYC, transaction confirmation, or the optional reversal phase (Table \ref{table:tab1workaroundsContinuum}). In the following sections, we discuss these workarounds and their motivations (Table \ref{table:tab2workaroundmotivations}). Workarounds can additionally (roughly) be categorized by what component of the MoMo environment they modify; \textit{the user}, \textit{the agent}, or \textit{the transaction} itself.  

\begin {table} [htbp]
\caption{Workarounds through the transaction stages}
\centering
\begin {tabular} {|l|l|l|}
\hline
\textbf{Workaround}& \textbf{Ke (n)}& \textbf{Tz (n)}\\
 \hline
 \hline
\multicolumn{3}{|c|}{Transaction Execution Stage}\\
\hline
Proxy, remote and direct transactions& 23& 18\\
Leaving money with the agent& 18& 20\\
Modify transaction size and/or location& 10& 9\\
Advance transactions& 5& 10\\
System circumvention& 7& 6\\
Agent-enabled credit& 2& 5\\
\hline
\multicolumn{3}{|c|}{KYC Stage} \\
\hline
Relationship-based KYC & 20& 1\\
Reduced KYC using ID number only& 19& 0\\
Third-party ID use& 12& 0\\
Complaisant agents& 7& 0\\
\hline
\multicolumn{3}{|c|}{Transaction Confirmation Stage}\\
\hline
Reliance on agent confirmation & 4& 14\\
Get recipient's verbal confirmation& 1& 18\\
Alternative confirmation (Check balance)& & \\
\hline
\multicolumn{3}{|c|}{Post-transaction Execution} \\
\hline
Agent-assisted reversal & 7& 8\\
Multiple transaction  attempts& 1& 3\\
\hline
\multicolumn{3}{|c|}{\textbf{(n): number of respondents who mentioned}} \\
\hline
 \end{tabular}
\label{table:tab1workarounds}
\end{table}

\begin {table*} [htbp]
\caption{Top seven motivations for pursuing workarounds}
\centering
\begin {tabular} {|p{5cm}|l|l|p{9cm}|}
\hline
\textbf{Motivation}& \textbf{Ke (n)}& \textbf{Tz (n)}& \textbf{Sample Quote}\\
 \hline
 \hline
Convenience and time-saving& 22& 26&  ``I had to leave [the money]. I didn’t want to wait because I was in a hurry" (KE19) \\
\hline
Navigating MoMo costs&15 & 21& ``I was trying skip the transaction charges because when you put it into your phone you will be charged the fees to complete that transaction" (TZ20)  \\
\hline
Unavailable service (Network downtime and Insufficient float)& 10& 14& ``I went to an agent and he didn’t have float and so I left money and the [phone] number on a paper and deposited for me later." (TZ11) \\
\hline
Surveillance concerns& 16& 16& ``There is an issue with the withdrawal. I cannot go to someone [an agent] I know because they can send someone to follow me. If you want to withdraw such amounts, you do it in town where nobody knows you" (KE19) \\
\hline
Physical ID challenges& 18& 1& ``[The requirement to have and ID] is an issue because sometimes you forget" (KE31) \\
\hline
Navigating KYC denial of service& 15& 2& ``I do not have an ID, they[the agent] know I do not have an ID. They do not ask me a lot of questions. If the agent sees me, they just deposit the money without asking" (KE32) \\
\hline
Usability issues (Data entry, complex interfaces etc)&15 & 16& ``I was withdrawing money but unfortunately I entered the wrong agent number and so the money went to another agent." (TZ24) \\
\hline
 \end{tabular}
\label{table:tab2workaroundmotivations}
\end{table*}

\subsubsection{Considerations in choice of agents}
While factors like distance from their home, float, and size of transaction were important considerations in choosing an agent, users also considered the perceived security based on the agent's physical location (n=23), and indicated that they preferred to use specific agents regularly (n=58). The top reason for this preference was trust (n=47).  For most, this trust stemmed from knowing the agent and perceiving them as honest. 
``\textit{If you know each other, that means you have built the trust}" (TZ13). 
Other reasons for trust were characterized by a utilitarian perspective---what the user could do with their agent---and were similar to the other reasons for using specific agents regularly like easier recourse, traceability, being assured of float, informal transaction arrangements and perceived confidentiality by the agent (Table \ref{table:regularagents}).
Overall, we observed that users choose agents largely based on security and/or convenience. Unlike the study by Chamboko et al, \cite{chamboko2021role} we did not observe any major considerations around gender. When participants mentioned a preference for agents of a particular gender (n=10) it was generally due to their sense of which agents were more personable.

\begin {table*} [htbp]
\caption{Reasons why users use agents regularly}
\centering
\begin {tabular} {|p{2.7cm}|p{1.5cm}|p{10.8cm}|}
\hline
\textbf{Reason}& \textbf{(n)}& \textbf{Excerpt}\\
\hline
Trust& 47& ``When we are dealing with something like money, the word trust has to be there." (KE07)\\
\hline
Being known& 34& ``When you call them at that time or any other time, they'll just accept your request because they already know that you are their customer" (KE02)\\
\hline
Informal agreements& 25& ``If I have a money issue and I do not have money, at times they lend me." (KE20)

\medskip
``Because sometimes I have to withdraw funds when I am not present to give to someone who is present to keep things going" (TZ12)\\
\hline
Easier recourse& 12& ``Even if I made a mistake, that means I can go back to him and its easy for him to understand me." (TZ11)\\
\hline
Traceability/permanence& 10& ``I know where to find them. Yeah, like not necessarily at their work, [but also] their home" (KE08)

\medskip
``Other agents might do something bad and when you come back the next time you won't find them... Because they just have a small place arranged for this particular service, and they might even move elsewhere. I trust those ones that I can go and buy products from them as well...somebody with a shop and he is doing other transactions" (TZ27)\\
\hline
Easier KYC& 7& ``Because I do not have an ID for this place [Kenyan ID]. So I will go to the one who is further because we know each other." (KE24)\\
\hline
Loyalty& 7& ``There is an agent I prefer because they are faithful to you and you are faithful to them" (KE25)\\
\hline
Assured of float& 6& ``I prefer them because they are reliable with their work. Others would tell you to leave your number to send it later. That is what I don’t like." (TZ31)\\
\hline
Confidentiality& 5& ``The agent is supposed to be someone who can keep secrets because you don’t know what intentions one might have with you if they share information about your transactions" (TZ16)\\
\hline
\multicolumn{3}{|c|}{\textbf{*(n): number of respondents who mentioned}}\\
\hline
\end{tabular}
\label{table:regularagents}
\end{table*}

\subsubsection{Workarounds in the transaction execution phase}
% \mbox{}
% \medskip
Among workarounds that happen during transaction execution, we categorize them as follows: (1) Alternative transaction executions in which either the agent or client does not execute their intended role in the transaction, %(Fig. \ref{figDeposit} and \ref{figWithdraw}), 
(2) agents provide alternative services that are not endorsed by MoMo, and (3) clients modify the transaction characteristics (e.g., size, location).
\mypara{\textbf{Alternative execution with changes in agent/client role}}
\label{lbl:alternativeTxExec}
This first category of workarounds at the transaction execution phase entailed users working with agents to navigate various challenges by having agents either complete the transaction on their behalf, or having them act as transaction proxies. For example, to manage their transaction charges, users completed transactions through direct deposits. In this case the agent directly deposited into the recipient's account (Avoiding steps 1-3 in Figure \ref{P2P}). TZ27 said,  ``\textit{When [the agent] deposits it into my account and I send it from my account, I will be charged, and that’s why sometimes I just decide to ask the agent to do the transaction straight away.}" 
In addition to saving on cost, this workaround transferred to the agent the responsibility for potential mistakes. The transferred responsibility was motivated by the complexities of the reversal process, which we discuss further in Section \ref{workaroundpostexec}. 

Some users also reported leaving money with the agent to execute the transaction later. This was mostly motivated by convenience either because they were in a hurry and/or the MoMo network was unavailable, or when the agent did not have sufficient float (Section \ref{sec:floatdef}). ``\textit{If there is a delay... I will just have to leave the money with my details and phone number}” (KE01). Other times, insufficient float necessitated partial fulfillment of transactions where the user collected part of the cash and got the rest later. Because of the burden of liquidity management, which falls squarely on agents, \cite{eijkman2010bridges} leaving money or getting money later benefits the agent as well, but may be problematic too when users cannot cash out their money \cite{kenya2009mobile}. Some of those who experienced difficulty in using MoMo reported having an agent completing a transaction on their behalf.

For P2P transactions, agents served as proxies as users made them cash collection points for the intended recipient. In addition to being useful when a MoMo user wanted to transact with a non-registered individual, or when the recipient did not have the required identification---such as when using a SIM card registered under another person, agents acting as proxies was also motivated by the need to save on transaction charges. This workaround circumvented steps 2-4 in Figure \ref{P2P}, by allowing the user to cash out at recipient's agent instead. “\textit{Maybe you have a friend, who needs like a hundred shillings, instead of sending money to the friend, to avoid that transaction cost, you just withdraw in a certain agent then you tell your friend to go and collect}” (KE17).

While the intended design of MoMo involved transacting in person to facilitate KYC, users modified this by sending a third party to complete the transaction on their behalf. Although this sometimes entailed sending the person to make a deposit, more participants indicated that they would send the proxy to collect physical cash once they had remotely initiated the cash-out. Sometimes, the individual who was sent was the intended final recipient. Many participants stated that they completed transactions without being physically present, mostly for convenience's sake.
\mypara{\textit{Privacy and security issues in alternative execution}}Using direct transactions means that the agent received all the transaction notifications and that the user had no means of recourse in the event of a problem as they had no proof of the completed transaction. The alternative execution workarounds also exacerbated the need to share more personal information with both agents and other parties acting as proxies. For example, in direct transactions, Alice has to share with her agent Bob's personal details, and when leaving money with the agent to transact later, users reported leaving their personal details such as their ID, phone numbers, and names with the agent. These were often written ``on a piece of paper.'' Alternatively, users gave a proxy some or all of the following details to share with the agent for purposes of verifying the transaction: their names, ID and phone number, amount transacted, and the transaction confirmation message that they received from the MoMo provider. Remote and proxy transactions therefore thwarted the KYC measures put in place by MoMo providers that require the customer to show up in person at the agent to facilitate identification and authentication (Figure \ref{figDeposit} and \ref{figWithdraw}). As a result, some participants reported that the agent kept their personal details such as their phone number and ID number to facilitate such transactions when they were not physically present. This is similar to a previous finding on cybercafe managers keeping their patrons' passwords \cite{munyendoeighty}.
\mypara{\textbf{Alternative services: agent loans and advance transactions}}
For this second category of workarounds at the transaction execution phase, the findings show that in addition to facilitating CICO services, agents offered loans and advance transactions to users. These interactions were not accompanied by any formal agreements. TZ07 indicated that he gets a loan from the agent to settle employee wages when ``\textit{working at the site and I need to pay the workers and I have shortage of money.}" KE09 even reported that she can get ``\textit{a certain amount of money, and return it at the end of the month.}"

On the other hand, advance transactions allowed the agent to disburse e-money or physical cash before the actual MoMo transaction. ``\textit{When you are in a hurry, you can even go to the agent and take some cash, then withdraw later}" (KE04). KE16 also said that she would call the agent and ``\textit{tell him to deposit some money for me. `When I come back, I will give it to you'.}"  Instead of getting cash, TZ02 shared that they can get an airtime advance.  ``\textit{I can go and tell her to send an airtime bundle to me, she does so and then I will give her the money later.}"
\mypara{\textbf{Modifying transaction characteristics}}
\label{sec:TxModification}
In this third category of workarounds, depending on what challenge they were trying to address, users modified transaction characteristics altering the size, location, or channel of transaction execution. Users reported modifying the transaction size to save on transaction fees. For example, we observed that users in Kenya split their transaction to smaller payments that did not attract any charges, as opposed to sending the whole amount as a lump sum. ``\textit{[Say] I want to send you Ksh 300. If I send you like a hundred separately three times, that will save on cost because sending Ksh 100 is free}" (KE08). In Tanzania the providers imposed fees on all P2P transactions, even of the smallest value. However, we still observed a different way of saving on cost through the use of alternative transaction channels that circumvented the MoMo system. This was enabled by the use of person-to-business (P2B) payment channels instead of the normal withdrawal channels through the agent's store number. In this case, the consumer did not incur any charges---even for large amounts. ``\textit{if you want to withdraw money and you don’t want to be charged a lot they tell you to use \textit{LIPA Number}\footnote{LIPA number is a merchant-specific number that is reserved for settling payment for other goods and services that the agent may provide. The consumer pays no service charge.}}" (TZ18). Users also modified the transaction size and location for privacy and security reasons, which we discuss separately below. 
\medskip
\mypara{{\textit{Privacy/security issues in modification of transaction characteristics}}} Many users were apprehensive about agents knowing their transactions details. ``\textit{I don’t feel good because you can’t know someone’s intentions. [The agent] can know and tell someone else and then you are followed up and robbed}" (TZ32). In prior work, users in Nigeria expressed similar sentiments, reporting the fear of being robbed at the point of cash withdrawal \cite{mogaji2022dark}. 
Participants therefore indicated that they sometimes changed how much they transacted and this afforded them some perceived security/privacy. For KE08, ``\textit{if the transaction is very large, I don’t have to transact it all at once. Maybe I can divide it into two, maybe three times. Maybe do one transaction here, then another in another agent like that. You can do that for security reasons.}" While many users alluded to some fear of robbery, others reported social concerns as a reason for desiring privacy from the agent. For example KE16 said, ``\textit{when you are transacting a small amount, there is shame [especially] when I am a male and I go to a female agent.}"

In addition to changing the transaction size, some users reported changing the location where they transacted to achieve more security/privacy. TZ33 shared that, ``\textit{If for instance this week I have withdrawn a large amount from [one agent], and if next week I am withdrawing almost the same amount of money then, I would not withdraw from [the same agent] this time.}" Some of those who reported feeling  embarrassed about sharing their transaction amount with the agent said that they would travel further to a place where they are not known.  TZ07 explained, ``\textit{I might feel shy to withdraw TShs 6,000\footnote{One US dollar was approximately equivalent to 2340 Tanzanian Shillings at the time of writing this paper.} and would say, `I don’t want my nearest agent to know'.  [So] I will go very far where no one knows about me and withdraw.}" This same participant brought up the benefit of ATMs when compared to transacting at an agent. ``\textit{If I go to withdraw 10,000 at the ATM, it is my secret, or 200,000, it's my secret.}"
\subsubsection{Workarounds in the KYC phase}
\label{Sec:KYCworkarounds}
\mbox{}
\\
KYC practices in Kenya and Tanzania were implemented differently at the time of our study.  In Kenya, MoMo users were required to show their physical ID to the agent whenever they transacted. Tanzania did not have similar requirements at the time of the study. We therefore skipped the questions addressing user perceptions about using  IDs to transact in Tanzania. Thus any sentiments that arose emerged organically as seen in (Table \ref{table:tab1workarounds}). When asked how they felt about the requirement for an ID to transact, KE06 said, ``\textit{It's good to provide IDs because most people are fraudsters.}"  Participants from Tanzania indicated that physical ID was a requirement in the past, but these checks happened rarely at the time of the study, depending on the size of the transaction and the agent. However, they too were concerned about the risks this might present: ``\textit{I am using a SIM card which is not mine... and nobody asks about it. What if I stole that SIM card? Therefore, [the ID] is very important}" \footnote{Up to 18\% of SIM cards in LMICs are registered under a third-party's ID, usually because the people have no ID of their own.}(TZ07). These responses suggest users' beliefs regarding the importance of KYC for security. TZ16 shared a negative experience stemming from lack of identification. ``\textit{I told you of how money was withdrawn from my account. If they required to be shown ID that would not have happened because the SIM card was registered with my name.}"

While users expressed beliefs about the importance of KYC, many of them also felt that KYC was ``cumbersome," ``frustrating," and ``not good, because they require the ID every time." Further analysis of the data revealed that KYC using a physical ID card presented multiple challenges that we summarized as being forgotten, lost, or simply not owning one. As a result, individuals adopted workarounds to navigate the physical ID challenges and the potential denial of service from falling short of requirements. We discuss the two categories of KYC workarounds that our findings surfaced.
\\
\mypara{\textbf{Agent-adjusted KYC}}
By transacting at agents they already used regularly, participants said they avoided  having to show an ID because the agent knew them. KE19 said, ``\textit{It's not a must if the agent knows you... I carry my ID when I want to withdraw money in a place where I am not known, maybe in town or places that are far from home.}" Some agents instead acted out of goodwill, or users looked for agents with lax KYC practices. KE06 urgently needed some money because they were taking their child to school ``...\textit{I went to an agent who refused to help with the transaction because I had no ID. I was upset...I went to another agent to whom I explained the situation to and told him/her that I had memorized the ID number. The agent asked me if I was sure the ID was mine about three times. I said yes...The agent allowed me to transact}." 
\mypara{\textbf{Alternative identity proofing}}
% \medskip
In place of the standard KYC procedures, users and agents sometimes adopted alternative identity proofing in which users: (a) gave their ID number instead of producing the physical ID card, (b) used a third-party ID, or (c) relied on third-party recommenders. KE08 remarked that,  ``\textit{...some agents don’t even bother asking for the physical ID. They just ask for the number.}” Some of those who did not use a SIM card registered in their own name presented the ID of the person under whom the SIM card was registered. ``\textit{I used my cousin’s ID because I had already registered the line with that ID, so I had to show his ID because the name that came was not my name}" (KE15).
When users could not present the third party's physical ID, they either provided the ID number or a copy of the  ID. ``\textit{I was using my friend's ID. I had his photocopy}" (KE16). Two participants from Kenya mentioned ``identifying themselves through other means" like ``\textit{a friend who can confirm that this so and so}" (KE12).
\mypara{{\textit{Privacy/security issues in KYC workarounds}}}
 In addition to agents keeping customer information for purposes of transacting in their absence, as we discussed earlier (\ref{lbl:alternativeTxExec}, the KYC workarounds modified what was provided and how it was provided and these were divergent from the KYC processes as stipulated. Agent-adjusted KYC and using a third-party ID changed what was provided. By using third-party recommenders and an ID number in place of the physical card, users and agents modified how KYC was achieved. The purpose of showing a physical ID in Kenya was to facilitate authentication by the agent, who compared the name on the ID with the name appearing on the agent's transaction summary. This was essentially the name under which the SIM card was registered. Lax KYC procedures by agents exacerbated the privacy and security risks arising from KYC workarounds. The findings show that not all agents implemented KYC in a standard manner. Whether an agent asks or does not ask you for an ID \textit{``depends on the place you usually go to deposit or withdraw money}" (KE05). For example \textit{``in case I am in Kayole,\footnote{Kayole is a low-income neighborhood in Nairobi, Kenya.} I have to have an ID, and when I am in Kibera,\footnote{Kibera is a low-income and largely informal settlement neighborhood in Nairobi, Kenya.} it’s not a must}" (KE20). MoMo users also felt that agents were not always vigilant with regards to implementing KYC. ``\textit{They are not keen}" (KE16). ``\textit{Sometimes I am a boy and I have gone with a girl’s ID}" (KE15) or ``\textit{they would ask `what is your Name?’ I tell them ‘XXX YYY.' They can see I am a male, but I mentioned a female name and nobody cares. Whether I am a thief, I stole her SIM card or whatever, nobody cares and it is actually very concerning}" (TZ07).

\subsubsection{Workarounds at transaction confirmation}
\label{sec:tx_confirmation_workarounds}
\mbox{}
\\
When asked about how they knew that a transaction was complete, the two most common ways users mentioned were the transaction confirmation message by the MoMo provider (n=61) and checking their balance (n=31). However, due to network downtime, ``\textit{...the message sometimes delayed}" (TZ31). Sometimes, only the balance checking option was available. Unfortunately in addition to challenges with timeliness of the message, these two methods were not always dependable. ``\textit{Some messages might be fake}" (TZ27) (We discuss these fraud-related concerns further in Section \ref{sec:fraud}) and ``\textit{to check balance, they charge some amount. It is not free}" (TZ12). As a result, users designed other transaction confirmation practices namely: reliance on agent confirmation and getting the recipient's verbal confirmation.
\mypara{\textbf{Reliance on agent for confirmation}} Users relied on agent communication in many instances. When they were transacting at a distance, some users depended on the agent to let them know ``\textit{if there is a problem.''}  KE05 explained, \textit{``...that’s why I prefer to go to one agent who is close}" (KE05). In remote and direct transactions where the user did not receive the transaction notifications, users waited until the agent told them the name that had appeared on their side of the transaction summary. ``\textit{When I give [the agent] the money to send to someone, when he mentions the recipient’s name and it matches the one that I am sending to, then I know it has gone through}" (TZ06). As an additional precaution, TZ24 said, ``\textit{what I usually do is that I don’t tell the agent the recipient’s name that will appear, I would wait until the agent tells me ‘is it so and so name?’}"
\mypara{\textbf{Getting recipient's verbal confirmation}}In addition to getting confirmation from the agent, some participants also called the recipient to confirm that they had received the money. This was especially common when the user asked the agent to directly top up the recipient's balance (see Section \ref{lbl:alternativeTxExec}). ``\textit{I would call the person I am intending to send money to before [to let them know], that I will be sending money through an agent and not through my number}" (TZ13). %"\textit{when the person who I am sending the money confirms 'Yes, I have seen [the money' then I leave}" (TZ01).
These multiple confirmation methods seemed to be necessary ``...\textit{because there are times that when you go to transfer money to someone else, the name comes up but when you ask the person whether they received it, they say ‘no, I have not yet received it'}" (TZ17).

\subsubsection{Workarounds at the optional transaction reversal phase}
\label{workaroundpostexec}
At the post-execution stage, the main workaround we observed was related to transaction reversal. The need for reversal mostly arose due to challenges in data-entry. The most common mistakes that users cited were entering the wrong agent number and sending to the wrong recipient. For such erroneous transactions, MoMo providers provided a self-serve reversal process. However, these reversal processes were not always clear. ``\textit{I had a challenge reversing it. So I went to the agent and asked if they could help.}" (KE1). They also did not always result in users getting back their money since success depended on the cooperation and honesty of the unintended recipient. If the latter had already withdrawn the money or transferred it out of their mobile wallet, a user could not get it back. Even in instances when the reversal process was successful, users did not get the money back immediately.``\textit{I had to wait for seven days}" (TZ09). In general, the cost of making a mistake was high for users as aptly indicated by KE15, ``\textit{I saw that when you make a mistake, you will lose money so you have to now be keen.}"

\subsection{Q2: Dealing with reduced privacy/security}
\label{sec:enablers}
As seen through the findings in Section \ref{sec:habitsandpractices}, the user workarounds were often accompanied by more risks and reduced security/privacy. As a result users took further measures to mitigate such risks. Based on user responses, we identified two categories of measures:  using a familiar agent, and measures to limit personal risks. 

\subsubsection{Using a familiar agent} 
Close to half of the participants (n=36) believed that the agent agreed to help because they were a customer. Assent by the agent was deemed necessary for ``courtesy" and ``building friendship" because some of these agents were neighbors: ``\textit{...we live close to each other}" (KE24), and because \textit{``If he helped me it’s easy to send other people to his place for transactions}" (TZ09).

While the perception that success was based on a ``customer being king'' philosophy, a majority of participants (n=51) mentioned that knowing the agent gave a sense of security. 
``\textit{I felt it was not good [to leave the agent the money] but, I just trusted anyway...because I usually use that same agent}" (KE01). This concept of knowing the agent as an enabler was also evident in the way users reported ``doing the safe thing'' when they did not know the agent. For example, with regards to waiting to see the transaction confirmation message, TZ09  said, ``\textit{If the agent is a stranger, I wait until I receive it.}"

\subsubsection{Measures to limit personal risk}Participants spoke about actions they took to limit personal risk. Most were similar to those already discussed as workarounds during transaction execution: i.e, changing the transaction size or location depending on the perceived risk.
TZ16 said, ``\textit{...I always look at the amount, I wouldn’t have left one million shillings [for the agent to deposit later] because I would have got sick, if it got lost.}" People who lacked technical literacy and relied on agent assistance mentioned withdrawing all the money from their wallet. While they could not navigate the interface to know their balance they could count the money to ensure it was the full sum of what the sender had indicated. In proxy and remote transactions, users and agents designed alternative ways of limiting risk including calling the agent ahead of the proxy's arrival to authorize them to disburse money to a proxy, or the agent would call the sender. Some users even went ahead and introduced their elected proxies to the agent, and this made authentication in future proxy transactions unnecessary. Users also limited their risk by involving proxies less in transactions that would required them to share information such as their MoMo PIN or to send the proxy with the phone. As a result, most users reported using proxies to collect the cash after initiating the transaction remotely. ``\textit{I sent my brother...because I was just making a deposit and not a withdrawal that would require me to give them the PIN. If I gave him the PIN maybe he can go and withdraw everything}" (KE25). Users also seemed to exercise caution in who they choose as a proxy. More often than not, the proxy was a trusted or known individual.

\subsection{Q3: Implications and challenges of user-agent interactions and  practices} 
\label{sec:risksandchallenges}
Some of the risks and challenges that arise from using MoMo relate to privacy and security. Others stem directly from workarounds, while others are a result of the current MoMo system structure (as noted by \cite{mogaji2022dark}).

\subsubsection{Unofficial and agent-determined fees}
When users worked with the agent to circumvent official transaction costs, they were often subject to ad-hoc agent-determined charges. ``\textit{You may find that he tells you “Please give me Tshs 500 because I have gained nothing from this transaction.” You have to do it because if you do not, next time you won’t be helped}" (TZ01). For some users, there was no other option because the lack of interoperability made the workaround necessary. ``\textit{When you go to send money in another mobile network, the agent asks you to add a certain amount.... They claim that to send money in a different network is a little bit costly}" (TZ34). This risk of being overcharged has been noted before \cite{martin2019mobile,mogaji2022dark}.  

\subsubsection{Reduced data privacy}
As highlighted in previous sections, the alternative ways of transacting such as proxy,  remote, and direct transactions resulted in more sharing of personal and private transaction data (often in unsafe ways such as writing on paper) with multiple parties including agents and proxies to facilitate the various workarounds. 

 \subsubsection{Concerns about disclosed data}We asked participants about how their data is recorded and processed and how they felt about giving personal details at the agents. Close to half (n=29) were concerned about disclosed data and indicated feeling ``insecure." 
 When we asked why they felt this way, most users cited the possibility of agent misconduct including sharing it with third parties like ``politicians" and ``random strangers." ``\textit{I think the agents are using those details wrongly. Maybe that’s why people receive scamming messages}" (TZ15). KE25 added,  ``\textit{...agents can use your ID to register other people.}" TZ27 had experienced the same, ``\textit{I discovered when I was going to register for my SIM cards that there were two other SIM cards that were registered under my ID}." Some of these concerns with agents have been noted before \cite{martin2019mobile}.
For others, the hesitation was purely a matter of privacy. Agents got to know the full names of their customers and some were not happy with this for reasons like not wanting to be called by their full name. 
\mypara{Willingness to share data} Despite these concerns, many participants were willing to share their data (n=45); half of these had previously indicated being concerned about disclosed data. The participants gave various reasons for why they shared their data. Most said that ``it was a requirement" for service access and they ``had no other way." There were those who felt that the information they left including the transaction data trails were important for security so that ``\textit{if an issue happens...a person cannot deny you because you have written [your details] there}" (TZ26).
In contrast to those who identified with the need for security, there are those who said they were not worried because ``the money is theirs" and they ``have not stolen it" or because they believed they were not sharing any sensitive information, where sensitive mostly referred to their PIN number. ``\textit{I get scared of sharing my PIN number but as for such other details like contact number and names, no, I don’t worry.}" (TZ06). The belief that agents were trustworthy and the belief that as users they were anonymous to the agent also made some users confident to give their information. 
``\textit{There is no problem when they know...because I withdraw and leave. They don’t know me}" (KE36).

\subsubsection{Risks from MoMo-related fraud}
\label{sec:fraud}
Some users were concerned about MoMo-related fraud. The types of fraud described by users  were similar to previously-documented fraud \cite{buku2017fraud} \cite{akomea2019control}; many of these were enabled by  social engineering, such as fake transaction reversals and masquerading as a customer care representative or agent. An example of the former is where fraudsters tricked the target user to believing they had sent money erroneously and asked them to send it back. Common reasons that the masquerading caller gave for the call involved resolving account anomalies. As the user complied by giving the required information, the fraudster was able to execute fraud. 
Participants also acknowledged that access to personal information as well as transaction trails facilitated fraud like SIM swaps and other forms of identity theft. For example, users in Kenya strongly associated identity theft frauds with the loss or unauthorized access of a person's ID. ``\textit{If you lose your ID, somebody can replace your SIM card, if that person knows your number, your last withdrawal, and balance, he might call customer care and give these details. After that they will withdraw, and later when you replace your line, you find there is nothing [in your wallet]}" (KE17). 

\section{Discussion}
\label {Sec:discussion}
% \subsection {Role of agents}
Our study joins several others in highlighting the centrality of agents to the MoMo ecosystem \cite{mogaji2022dark}, e.g., by ensuring the inclusion of users with limited technical capabilities. However, our study enriches this understanding by also illustrating the ways in which agents and users modify intended MoMo procedures (workarounds) to better meet users' needs, and the underlying causes for these workarounds.
In this section, we list some takeaway messages that emerged from the results in Section \ref{Sec:Results}. 
\mypara{Takeaway 1: User workarounds involving agents are often motivated by efforts to limit risk or uncertainty (e.g., due to environmental factors or user interface limitations)} Agents play a complex role in the MoMo ecosystem, mitigating risk in some cases, and introducing risk in others. For instance, agents regularly help users revert transactions sent to the wrong recipient, despite user interface tools meant to prevent such events. These challenges may be exacerbated in LMICs where users often use devices with small screens and that support technologies such as USSD and STK instead of apps. Users therefore have to manually type the recipient number instead of selecting from a recipient list as is common with mobile payment apps in the US. These challenges motivate some workarounds, such as people writing numbers on a paper (to enable easier transfer), or asking the agent to enter the number,
thus transferring responsibility for data entry risks to the agent. 
 
Treating agent interactions as a risk-management measure is plausible given the terms of service in most MoMo implementations, which seldom protect the user from losses related to the use of the service \cite{bowers2017regulators}. This is unlike the United States where users of mobile payment services like Venmo and Paypal are protected by regulations and policies that limit liability for loss \cite{bowers2017regulators}.

Although agents often mitigate risks for users, they can also introduce new risks. 
Agents regularly handle sensitive and diverse user data. Concerns about agent surveillance, agent maleficence in perpetrating fraud, and inappropriate sharing/commodification of user information are all issues raised by our study participants (see next takeaway). 
\mypara{Takeaway 2: Users' stated preferences for security measures are inconsistent with their (workaround) behavior}

Our findings suggest that people are largely aware of  security threats related to the use of MoMo, and view security measures (like KYC and not sharing one's PIN) as important. 
Some of this awareness stems from campaigns by MNOs that caution users to keep their PINs secret \cite{mckee2015doing}. On the flip side, users often perceive themselves as safe as long as they have not shared their PIN; other personal data is often viewed as non-security-sensitive. While several study subjects detailed the importance of process security, e.g., for curbing fraud, they also engaged in contradictory behaviour, especially with regards to KYC. 
We found prevalent use of third-party IDs, as well as other workarounds to the expected KYC process. 
We hypothesize that this is a result of burdensome KYC practices that does not appear to meet the expectations and needs of the target population of MoMo users.   
\mypara{Takeaway 3: Users have varied perceptions on the role and impact of data sharing with agents} A significant number of users in our study reported consented data sharing,  mainly because giving personal information was a mandatory prerequisite for MoMo service access. Users also reported concerns over third-party information sharing although they had no way of ascertaining that this was actually happening. This is drastically different from many countries with data use restrictions; for instance,  US banks and mobile payment services are required by the Federal Deposit Insurance Corporation (FDIC) to have privacy notices where users can opt out of data collection and sharing \cite{bowers2017regulators}. 
These regulations also mandate that such services clearly state the purpose for collecting  user data and third-party data sharing. To the contrary, most MoMo services do not have privacy policies, and when they do, they are difficult for users to understand \cite{bowers2017regulators,munyendo2022desperate}, a challenge which has equally been noted in the US~\cite{schaub2015design}.

Other subjects were willing to freely share data with  agents because they ``had nothing to hide" or because they trusted the agent. 
Similar attitudes have been reported in prior work from other contexts, e.g., on the (lack of) need for privacy for `normal people' \cite{gaw2006secrecy} and the link between trust and willingness to share data   \cite{das2014effect,yisa2023investigating}.
Some felt that data sharing with agents was important to prevent fraud. Our study does not directly show evidence of risks from agents inappropriately sharing user data with third parties; however, we postulate that lax data sharing with agents could pose security vulnerabilities. 

Our observations regarding attitudes towards data sharing are particularly interesting in light of prior work (e.g., \cite{martin2019mobile,acquisti2015privacy}), which showed that the perception of being monitored plays an important role in people's behaviour. 
In fact, in Zambia, people migrated to MoMo from traditional banking services because of concerns about surveillance posed by their Taxpayer Identification Number \cite{phiri_2018}.  The bottom line, therefore, is for governments, regulators and technology providers to balance between protecting user privacy while having visibility. However, since governments can also misuse visibility to suppress and subdue citizens, this recommendation should be adopted with caution within a strong policy and regulatory environment that has the necessary checks and balances in place. 

\subsection {Remaining barriers to financial inclusion}
While MoMo has inarguably extended access and reach of financial services, 
the workarounds documented in this work suggest that there is still room for improvement.

Access continues to depend heavily on local and regional regulations. For almost a decade, most African countries have continued to mandate SIM card registration using some form of identification like biometrics or national identification numbers \cite{privacyinternational_2019} \cite{jentzsch2012implications} \cite{donovan2014therise}. Access to such ID is known to be challenging in many locations and contexts \cite{gsmmobile} \cite{clark2022id4d} \cite{demirgucc2022global} and we discuss some of these challenges in Section \ref{Sec:KYCworkarounds}. Once users gain access to MoMo, the cost-related workarounds documented in this work suggest that affordability may still be a barrier for many. 
This is expected to worsen as governments increasingly impose e-levies on MoMo transactions; 
e.g., MoMo use in many African countries dropped when  governments introduced more taxes \cite{clifford2020causes}. 

Finally, our study also suggests that usability challenges pose a barrier to the flexible  use of MoMo (Sections \ref{workaroundpostexec} and \ref{sec:tx_confirmation_workarounds}), and these challenges go hand-in-hand with  \emph{systemic} float- and network-related limitations (Sections \ref{lbl:alternativeTxExec} and \ref{workaroundpostexec}). 
For instance, while money transfer apps in the US allow for interoperability across banks, interoperability of MoMo services in Africa is still nascent \cite{naji2020tracking}. Some users cited these challenges as motivations for workarounds like direct withdrawals (Section \ref{lbl:alternativeTxExec}) which allowed the recipient to collect cash. Other times, agents who were interoperable (i.e, providing MoMo services for multiple MoMo providers) offered the transfer services through direct deposits but charged the user extra fees for offering this. Such barriers may reinforce reliance on cash. 

\section {Recommendations and Future Directions}
\label{sec:summary}
Nan et al., \cite{nan2021we} called for qualitative studies to better uncover the less understood themes of MoMo. 
Our study offers an in-depth understanding of the MoMo user-agent interaction which has largely remained anecdotal to date. Our findings show that users and agents work together to design alternative workflows, or workarounds. We offer insights on the challenges that users face, thus motivating these workarounds. By understanding and addressing these challenges, MoMo can be a more effective tool for digital financial inclusion. 
While we make no claims of generalizability, we hypothesize that the insights gained may be relevant to other African countries, since MoMo operations and regulatory environments have more similarities than differences \cite{bowers2017regulators, nan2021we}. 
Investigating whether MoMo usage patterns hold more broadly in Africa is an interesting question for future work. We provide some recommendations. 
 \\
 \noindent \textbf{Recommendation 1: Study how to improve the MoMo user experience for reverting and confirming transactions.}
 Our findings suggest that many users in both Kenya and Tanzania rely on agents (and/or workarounds) to revert or confirm transactions.
 This is despite the fact that in Kenya, the dominant MoMo's (M-PESA's) user interface includes an option to revert transactions within 25 seconds after a transaction is placed.
Hence, research is needed first to understand why users struggle with transaction reversal. 
Similar questions arise for transaction confirmation, though our study subjects generally attributed failures to receive transaction confirmation to network outages.
A better understanding of both problems, would pave way for the design of new interfaces that are compatible with mobile devices common in LMICs (e.g., feature phones) to improve this aspect of the user experience. 

\noindent \textbf{Recommendation 2:  Measure the effects of user privacy and security concerns regarding data sharing with agents}. 
Privacy and security concerns were a strong undercurrent in our results, with many users reluctant to share data with agents for various reasons.
We envision that semi-automated solutions could help address this issue, and reduce the prevalence of several workarounds. 
However, to know whether such solutions are worth the investment, we hypothesize that a deeper understanding is needed of user data sharing concerns with MoMo agents. 
For example, we would like to understand (a) what fraction of users would prefer to share less data with agents, and (b) what fraction of users actively change their MoMo behaviors (e.g. by breaking up transactions into smaller ones) as a result of privacy or security concerns with regards to agents. 

\noindent \textbf{Recommendation 3: Develop registration and KYC mechanisms for users who lack formal ID}. 
Stricter KYC requirements (both for onboarding and during cash-in or cash-out transactions) do not necessarily improve system security. 
They can instead lead to workarounds such as third-party SIM cards, especially for the many who lack formal ID. 
Although the root problem is lack of access to ID (a problem that should be a high priority to address, but will likely require changes to laws and government processes), we conjecture that as a stopgap measure, there is value to designing mechanisms that allow users to access MoMo without formal ID. For example, this could include biometric-based registration. 
Clearly, such methods will present tradeoffs (e.g., privacy/surveillance, operating costs), which should be explored systematically.

\section*{Acknowledgments}
The authors gratefully thank Michael Bridges for his valuable inputs on the design of this study, and our anonymous reviewers and shepherd for their constructive comments and suggestions. We also thank Leora Klapper and Rafael Mazer for their feedback. We thank all our colleagues and student assistants who helped with translation of the study tools. This study was made possible by the generous support of the Bill \& Melinda Gates Foundation. The views and opinions expressed in this study, however, are those of the authors and do not necessarily reflect the views or positions of the sponsors. 

\normalem
\bibliographystyle{plain}
\bibliography{references}

%\vfill\null
%\columnbreak
\appendices
\small
\setcounter{table}{0}
\renewcommand{\thetable}{A\arabic{table}}
\renewcommand{\thesection}{\Alph{section}.\arabic{section}}
\setcounter{section}{0}

\begin{appendices}

\begin {table} [htb!]
\caption{Study Approvals}
%\centering
\begin {tabular}{|p{1cm}|p{2.7cm}|p{3.3cm}|}
%\begin {tabular}[htb!] {|p{1cm}|p{2.7cm}|p{3.3cm}|}
\hline
\textbf{Country}& \textbf{Ethics Approval}& \textbf{Research Permits}\\
\hline
Kenya& Amref Ethics Review Scientific Committee& National Council of Science, Technology and Innovation (NACOSTI)\\
\hline
Tanzania& COSTECH& N/A\\
\hline
\end{tabular}
\label{appendix:approvals}
\end{table}

\section{Qualitative Codes}
\label{Appendix_Codes}
Values in parentheses correspond to number of people who mentioned each theme.
\begin{itemize}
    \item \textbf{Agent Choice}
    agent KYC practices (3), agent transaction costs (10), agent-level interoperability (7), distance matters (45), float and size of transaction (32), gender matters (11), good quality of service (23), location and security (23), prefer specific agents regularly (58)
    \medskip
    \item \textbf{Reason for Specific Agents}
    agent traceability and permanence (10), trust (47), assured of float (6), easier KYC (7), easier recourse (13), known (to each other) (34), loyalty (7), informal banking arrangements (25), maintain confidentiality (5)
    \medskip
    \item \textbf{Meaning of trust}
    assured of float (2), good customer care (7), honest (31), known (to each other)(14), informal banking arrangements (8), maintain confidentiality (4)
    \medskip
    \item \textbf{Workarounds}
    leaving money with agent (39), advance transactions (15), agent enabled credit (7), assisted reversal (15), modify transaction size (17), partial transaction fulfillment (4), proxy,remote and direct transactions (42), system circumvention (13), redacted identity proofing (19), relationship-based KYC (21), agent complaisant (7), third-party ID (12), third-party recommenders (2), agent confirmation (20), recipient confirmation (20), remote and proxy KYC (agent knows client information (5), phone authorization between client and agent (12), proxy bears required information (25), proxy known to agent (5))
    \medskip
    \item \textbf{Motivations for workarounds}
    surveillance concerns (32), convenience (49), cost avoidance (39), evade/delay digital loan deductions (2), interoperability issues (3), navigating user barriers (4),  service unavailable (25), Usability issues(23): (data entry (15),interface complexity (9), reversal complexities (7)), navigating KYC denial of service (17), physical ID challenges (19)
    \medskip
    \item \textbf{Workaround risks/challenges}
    agent determined fees (15), agent maleficence (14), delays and multiple visits (31), paper trail of personal data (12)
    \medskip
     \item \textbf{Why workaround is possible/successful}
    known agent (51), change behaviour on unknown agent(9), customer service by agent (35), institutional safeguards (3), personal security precautions [proxy is trusted (28), consider size of transaction (11)]
    \medskip
    \item \textbf{Data perceptions, practices and attitudes}
    concerns on disclosed data (29), perceived security (data is  non-sensitive (8), PIN not shared (22)), consented data sharing (agent is trusted (14), anonymity to agent (2), data trails for security (11), legitimate ownership (6), required for service (33)) 
    \medskip
    \item \textbf{Perceptions on KYC}
    cumbersome (15), necessary for security (24), should be flexible (4)
    \medskip
    \item \textbf{Fraud experiences}
    fake opportunities (5), fake reversal (10), identity theft+sim swaps(14), masquerading agent/provider (13)
     
\end{itemize}
\newpage
\section{Interview Guide}
\label{Appendix_IDI}
\textbf{General practices and agent choice}
\begin{enumerate}
    \item How frequently do you use mobile money?
    %\begin{enumerate}[a]
    \begin{enumerate}[label=(\alph*)]
        \item What do you use mobile money for?
        \item What do you use mobile money agents for?
        \item Would you have another way of completing all or some of these transactions without using an agent? If yes, which ones and how? 
    \end{enumerate}

\textit{I will read out some factors that may influence where people choose to transact. Let me know if you consider any of them by saying ‘Yes’ or ‘No’, and we will discuss them further: Distance to the agent? Gender of the agent? Being known to the agent?}

    \item (If distance is a factor) Under normal circumstances, how far do you travel to access an agent?
    \begin{enumerate}
        \item What circumstances (if any) could result in you traveling to an agent who is further? 
        \item If you had a large transaction, what would be your top considerations in choosing an agent and why?
        \item Do you always travel to this agent when you need to transact? If yes, why? If no, how do you transact then?
    \end{enumerate}

    \item (If gender is a factor) What is your preferred gender to transact with?
    \begin{enumerate}
        \item Why is the gender of the agent an important consideration for you? 
        \item What do you do when your more preferred agent (by gender) is not available? 
    \end{enumerate}

    \item (If being known is a factor)Do you have an agent or some agents that you prefer to regularly transact with?
    \begin{enumerate}
        \item Why do you prefer to transact at this/these agents regularly?
        \item (If trust is not mentioned) Would you say that you trust these agents? If yes, what made you feel that you could trust the agent(s)? If no, how did you arrive at the decision to choose these agents over other agents?
        \item What do you do when your preferred known agent is not available to transact?
    \end{enumerate}

    \item Are there any other factors that you consider in your choice of agent? If yes, please explain.
    
    \item Do you ever seek assistance from the agent? If yes - 
    \begin{enumerate}
        \item What kind of assistance do you seek from an agent to use mobile money?
        \item How do you decide which agent to visit when you need help and why
        \item Have you always received the assistance you needed from the agent? If yes, what do you think contributes to being able to get assistance every time? If no, Let's talk about one such instance when you did not receive assistance:
        \item what did the agent do or not do that made you feel unassisted?
        \item In your view, why do you think the agent did not help you?
        \item Did not getting help affect your mobile money decisions in any way?
    \end{enumerate}

    \item Do you ever get money from the agent to count money? If yes, What's the reason why you get help?
    \begin{enumerate}
        \item How do you go about getting the help you need: when you are withdrawing? when you are depositing? 
        \item How do you feel about getting help from the agent? Why do you feel this way?
        \item Have you ever had any trouble with an agent where they gave or deposited inaccurate amounts? If yes, what happened? 
    \end{enumerate}

\textbf{User transaction practices}
    \item Have you ever successfully completed a transaction that would normally require you to be physically present at the agent without being present? If yes
    \begin{enumerate}
        \item What kind of transaction were you completing?
        \item What circumstances made you complete the transaction this way
        \item How did you manage to complete the transaction without going to the agent
        \medskip
        \\
        \textit{If any part of the answer involves sending someone}
        \item Whom did you send and why did you choose this specific person? Did you have to send your phone, ID or PIN with the person?
        \item What actions did the person that you sent need to do to complete the transaction?
        \item What actions did the agent have to complete to fulfill the transaction? (probe: Did this need the agent to have your PIN?). If yes, how did you feel about the agent having this information? (If response suggests concern (why did you proceed anyway?)
        \item How often is it that you complete a transaction this way? Why?
        \item Have you ever tried to send someone to transact on your behalf and they were unable to complete it? If Yes - why were they unable to complete the transaction? 
    \end{enumerate}

    \item Have you ever left money with an agent and asked them to deposit it for you later? If no, why not? If yes:
    \begin{enumerate}
        \item What circumstances necessitated this? 
        \item How did you feel about leaving the money for the agent to complete the transaction?
        \item (For response given), what made  you feel that way?
        \item Why did you do it anyway?
    \end{enumerate}

    \item Have you ever presented cash to an agent with the request to deposit the money into someone else’s mobile money account? If yes, 
    \begin{enumerate}
        \item Why did you opt for this approach rather than sending it through your own number?
        \item Did the agent agree to send the money on your behalf? If yes, why do you think the agent agreed?
        \item Is there an occasion when the agent did not agree? If yes, why do you think the agent refused?
    \end{enumerate}

     \item In addition to the normal transaction charges that the MM provider, have you ever had to pay the agent extra fees for offering some service?
     \begin{enumerate}
        \item What service did you have to pay extra for?
        \item To the best of your knowledge, why do you think the agent charged you?
        \item How did you know that the agent had charged you extra fees? How did you feel about being charge by the agent?
    \end{enumerate}
   \item Has the agent ever extend additional services beyond make it possible for you to withdraw and to deposit? 
    \begin{enumerate}
        \item What favor did you receive?
        \item In your opinion, why do you think the agent extended this favor?
    \end{enumerate}

    \item Did you ever seek such a favor and the agent refused to help? If yes, why do you think the agent did not agree to help you?
    
    \textbf{Practices around transaction confirmation}
    \item How do you usually know whether or not the transaction went through?
    \begin{enumerate}
        \item At what point do you do you do this? [mention the confirmation that the respondent identified ]
        \item Is this what you always do, or are there times when you do something different? 
        \item When you do something different, what would you do and why?
        \item What do you do when there is a delay in the network?
    \end{enumerate}

    \textbf{KYC practices with ID}
    \item For which types of mobile money transactions are you required to show your ID?
    \begin{enumerate}
        \item Do you always carry your ID to complete transactions at an agent? 
        \item If No, when do you carry it? How do you decide when to carry it and when  not to? 
        \item Are there agents with whom you can regularly transact even when you do not have and ID? What makes it possible to transact at these agents?
        \item Have you ever found yourself without an ID at an agent? If yes – how did it go?
        \item How do you feel about the requirement to show IDs when transacting at agents?
    \end{enumerate}
    
    \textbf{Perceptions on privacy and security}
    \item How do you feel about the agent knowing your your transaction details such as the amount of money you have withdrawn or deposited? Why do you feel this way?

    \item How is your personal information (such as your name, phone number and ID number) received, recorded, and processed when you go to transact at an agent?
    \begin{enumerate}
        \item How do you feel about providing this information to the agent? Why?
        \item Does the way you feel influence any of your actions whenever you supply this information? If yes, how?
    \end{enumerate}

    \item (For those who require IDs to transact) have you ever lost or temporarily misplaced your ID?
    \begin{enumerate}
        \item Did this loss in any way affect your use of mobile money services? If yes, how?
        \item If no, what made it possible to continue using the services
    \end{enumerate}
    \item Have you heard of any mobile money-related risks that result from misplaced, fake, or stolen IDs?  If Yes – What are some of the risks that you have heard about?
    \item Have you ever needed to use someone else’s ID to transact at an agent. If yes, whose ID were you using and what circumstances necessitated this?
    \begin{enumerate}
        \item Was the agent aware that the ID was not yours? If yes, why do you think they allowed you to transact? 
    \end{enumerate}

    \item Have you or anyone you know experienced mobile money fraud or being cheated by agents or people pretending to be agents? If yes, who was defrauded and how did it happen?
    \begin{enumerate}
        \item (If the respondent says they have have been the victims) What happened?
        \item How did this incident affect you?
        \item Which people do you think are more susceptible to being defrauded and why?
        \item what circumstances do you think make people susceptible to fraud?
    \end{enumerate}

    \item Have you ever had any negative experience with regards to the use or misuse of your information (including breached trust by an agent)? If yes - 
    \begin{enumerate}
        \item Please recount to me any one such incident.
        \item How did you find out about the problem?
        \item (If they suggest the agent was responsible) What makes you believe the agent was responsible?
        \item How did this experience affect you?
        \item Did you do anything to resolve this problem? If yes, what did you do? If not, why not?
    \end{enumerate}
    
    \item have your experiences or awareness with/of mobile money fraud affected how or where you transact? If yes, how?
    \begin{enumerate}
        \item Has is affected how often you use mobile money? If yes, how?
        \item Is there anything you would do differently in how you use mobile money and interact with agents following this interview? 
    \end{enumerate} 
    
\end{enumerate}

\section{Theme Emergence}
\label{Appendix_ThemeEmergence}
\begin{figure}[!ht]
\includegraphics[width=\linewidth]{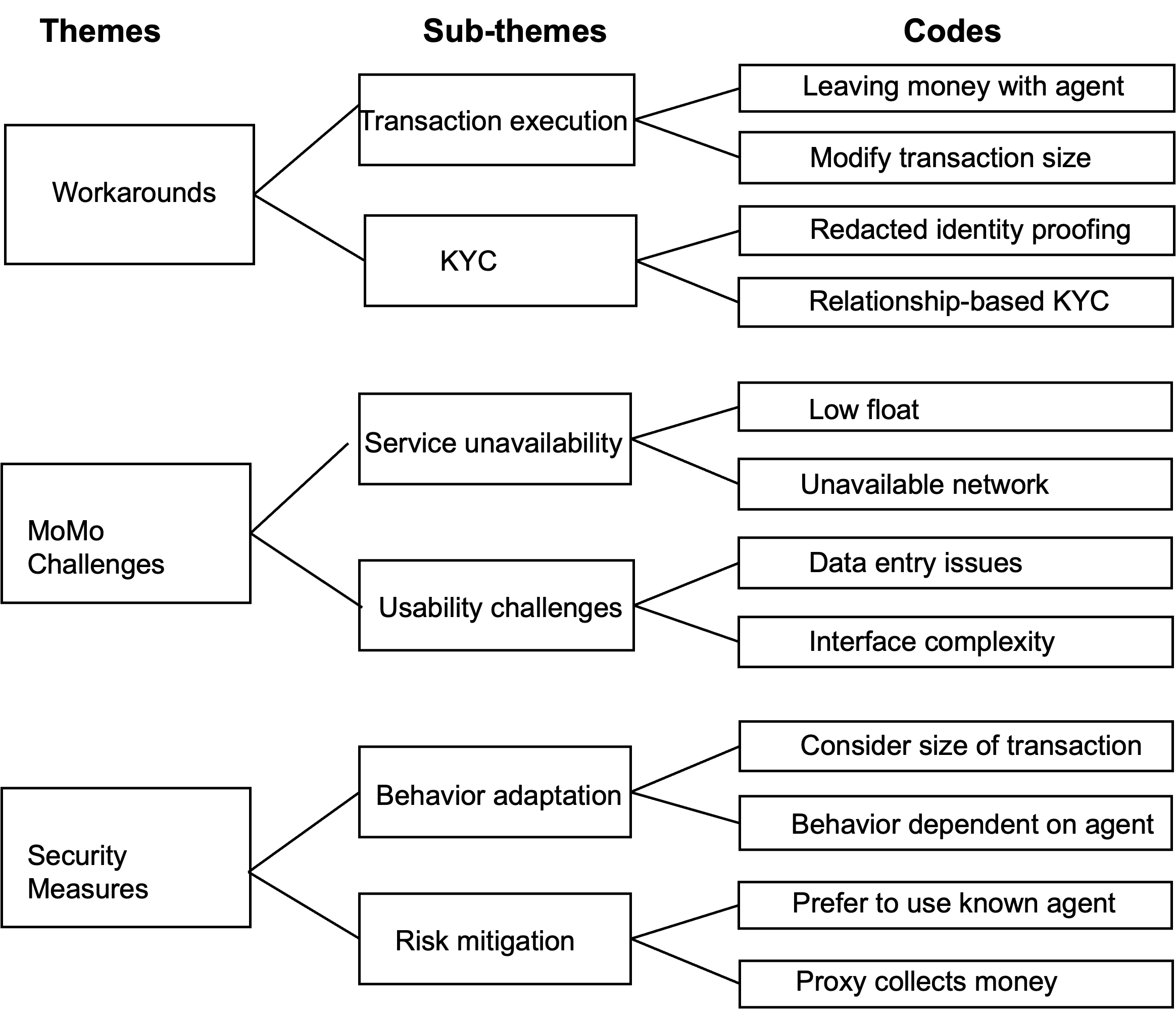}
\label{ThemeEmergence}
\end{figure}
\end{appendices}

\end{document}